# Apparatus for *Operando* X-ray Diffraction of Fuel Electrodes in High Temperature Solid Oxide Electrochemical Cells

*Jesse D. Benck, Daniel Rettenwander, Ariel Jackson, David Young, Yet-Ming Chiang**

Department of Materials Science and Engineering, Massachusetts Institute of Technology.
77 Massachusetts Ave, Cambridge, MA 02139.
*Author to whom correspondence should be addressed

## Abstract

Characterizing electrochemical energy conversion devices during operation is an important strategy for correlating device performance with the properties of cell materials under real operating conditions. While *operando* characterization has been used extensively for low temperature electrochemical cells, these techniques remain challenging for solid oxide electrochemical cells due to the high temperatures and reactive gas atmospheres these cells require. *Operando* X-ray diffraction measurements of solid oxide electrochemical cells could detect changes in the crystal structure of the cell materials, which can be useful for understanding degradation process that limit device lifetimes, but the experimental capability to perform *operando* X-ray diffraction on the fuel electrodes of these cells has not been demonstrated. Here we present the first experimental apparatus capable of performing X-ray diffraction measurements on the fuel electrodes of high temperature solid oxide electrochemical cells during operation under reducing gas atmospheres. We present data from an example experiment with a model solid oxide cell to demonstrate that this apparatus can collect X-ray diffraction spectra during electrochemical cell operation at high temperatures in humidified $H_2$ gas. Measurements performed using this apparatus can reveal new insights about solid oxide fuel cell and solid oxide electrolyzer cell degradation mechanisms to enable the design of durable, high performance devices.

# I. Introduction

High temperature solid state electrochemical cells such as solid oxide fuel cells (SOFCs) and solid oxide electrolyzer cells (SOECs) are promising technologies for efficiently converting between chemical and electrical energy [1]. A variety of SOFC and SOEC configurations can be utilized under different conditions, but all operate at elevated temperatures, typically within the range of 400 - 1,000 °C [1, 2]. The electrode and electrolyte materials that make up these cells can degrade or decompose during operation due to the combination of high temperatures, reactive gas atmospheres, and large electrical potentials [1]. As a result, widespread implementation of SOFC and SOEC technology has been limited by inadequate long-term durability [1].

While many previous studies have focused on degradation of the electrolyte or the oxygen electrode (the cathode in SOFCs or anode in SOECs), changes in the fuel electrode (the anode in SOFCs or cathode in SOECs) are also worthy of investigation because they can significantly impact cell performance [3, 4]. For example, in nickel oxide/yttria-stabilized zirconia (YSZ) anodes in metal-supported SOFCs, reduction of the nickel oxide to nickel can contribute to cell performance degradation [3]. New perovskites such as $Ln_{0.5}Sr_{0.5}Ti_{0.5}Mn_{0.5}O_{3\pm\delta}$ (*Ln*: La, Nd, and Sm) have promising properties that may make them attractive as a replacement for YSZ in SOFC anodes, but these materials may be subject to phase decomposition during operation [5]. Understanding the degradation processes that occur in these electrodes is critical to developing strategies that can improve the long term stability of solid oxide electrochemical cells [3].

*Operando* characterization techniques are becoming increasingly important for studying electrochemical systems because they enable measurements of composition, chemical state, and/or crystal structure during electrochemical cell operation [2]. These techniques have the ability to detect changes in the electrochemical cell over time and correlate these changes to operating conditions, which can yield important insights about reaction and degradation mechanisms [2]. While *operando* characterization has been employed extensively for low temperature electrochemical cells, the capability to perform these measurements on high temperature solid oxide cells remains relatively limited [2].

X-ray diffraction (XRD) is a powerful tool for measuring changes in the crystal structure of electrochemical cell materials. To date, there have been few previous reports of *operando* XRD measurements of high temperature solid oxide electrochemical cells [6, 7, 8, 9]. Some of these previous efforts have required synchrotron facilities [9], which can be difficult to access on a routine basis, limiting experimental throughput. Furthermore, the apparatuses used to perform these measurements were constructed for the study of oxygen electrodes and, therefore, consist of components that may not be appropriate for the reducing gas atmospheres such as humidified $H_2$ used for fuel electrodes [6, 7, 8, 9].

In this work, we describe the first apparatus that can enable collection of time-resolved XRD spectra of fuel electrodes in solid oxide electrochemical cells during electrochemical operation at high temperatures (up to 745 °C). The apparatus can operate under reducing (e.g. dry or humidified $H_2$) or inert (e.g. Ar or $N_2$) gas atmospheres, enabling measurements of fuel electrode materials, as well as electrolytes. It is designed to be compatible with a conventional laboratory powder X-ray diffractometer, which enables high experimental throughput. Setup of the electrochemical cell for testing is simple and requires no permanent bonds to make electrical contacts. Both two-electrode and three-electrode electrochemical cells in a range of shapes and sizes can be accommodated. This apparatus will enable measurements of structural changes in the fuel electrodes and electrolytes of solid oxide cells during operation, which can provide new insights into the design of high performance, durable solid oxide electrochemical cells.

## II. Design Objectives

Our apparatus is designed to enable collection of XRD spectra of the fuel electrode and electrolyte during solid oxide electrochemical cell operation. The primary design objectives are presented in Table 1. The apparatus is designed to be compatible with a Rigaku SmartLab X-ray diffractometer with a Cu K-α X-ray source, but could also be used with other X-ray diffractometers that have a large enough goniometer clearance to accommodate the apparatus. The desired XRD 2θ scan range of 5 - 90° is chosen to enable measurements of the highest intensity reflections of most materials. The components of the apparatus must be resistant to degradation under inert or reducing gas atmospheres to enable study of the fuel electrodes in solid oxide cells, which commonly operate in a $H_2$ atmosphere. The operating temperature range is chosen to enable measurements of most SOFCs and SOECs at or near standard conditions. The apparatus must be able to make a minimum of two electrical connections for the anode and cathode in a two-electrode cell, and a maximum of four electrical connections for a three-electrode cell with two redundant contacts to the working electrode. Finally, the apparatus must support a range of electrochemical cell sizes typical of solid oxide cells used for laboratory research.

**Table 1.** Design objectives for *operando* XRD apparatus.

| Specification | Range |
| --- | --- |
| XRD 2θ Scan Range | 5 - 90° |
| Gas Atmosphere | Reducing (dry or humidified $H_2$) or inert ($N_2$ or Ar) |
| Temperature | 20 - 725 °C |
| # Electrical Connections | 2 - 4 |
| Electrochemical Cell Diameter | 4 - 30 mm |
| Electrochemical Cell Thickness | 0.1 - 3 mm |

The SOEC used as a test sample for our apparatus is pictured in Figure 1. This cell is not a typical SOFC or SOEC, but is chosen a simple model system to provide a straightforward demonstration of the capabilities of our apparatus. The electrolyte is $BaZr_{0.8}Ce_{0.1}Y_{0.1}O_3$ (BZCY), a solid oxide proton conductor [10, 11] synthesized by CoorsTek Membrane Sciences AS. A Pt anode and reference electrode and Pd cathode are deposited onto a polished BZCY electrolyte disk using DC magnetron sputtering through Kapton shadow masks, which define the electrode geometries. The cathode and anode materials act as catalysts that promote the hydrogen evolution and oxidation reactions supplying protons to the BZCY. The Pd cathode can undergo electrochemical H insertion from the electrolyte and/or chemical H insertion from gas phase $H_2$ to form $PdH_x$ [12, 13], a structural change which is detectable by XRD. This phase transformation serves as a proxy for structural changes associated with degradation processes that could occur in SOFC or SOEC fuel electrodes. The cell has a diameter of 20 mm and thickness of 1 mm. Our apparatus must enable simultaneous high temperature electrochemistry and XRD measurements of the cathode lattice parameter in this cell.

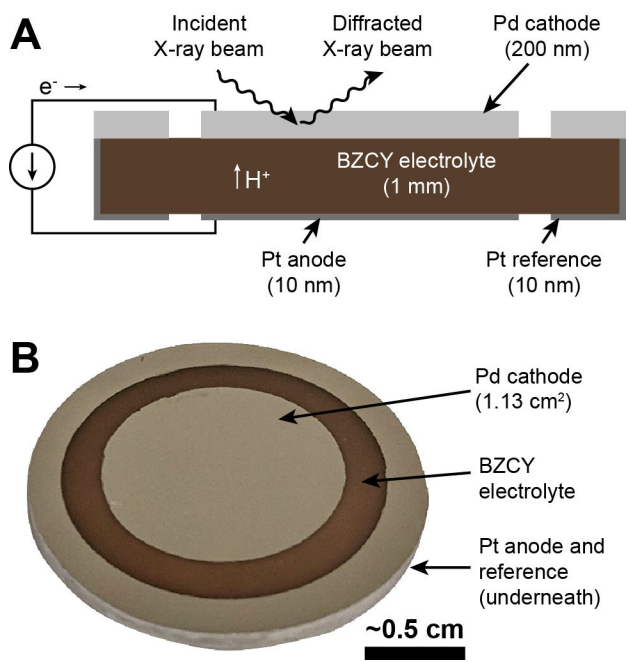

**Figure 1. A.** Side view diagram and **B.** photograph of the solid oxide electrolyzer cell used for our example *operando* X-ray diffraction measurements.

## III. Apparatus Design

Figure 2 displays photographs of the *operando* XRD apparatus. As pictured in panels A and B, the apparatus consists of a rectangular prism-shaped stainless steel housing with side lengths of 4 - 5 in. The bottom of the housing is welded to a 8 in outer diameter CF flange (Kurt J. Lesker Co.), which is opened only for apparatus construction or maintenance, and otherwise sealed with a silver-coated copper gasket during electrochemical cell setup and operation. The top of the housing is welded to a 4.5 in outer diameter CF flange, which is routinely opened to insert the electrochemical cell into the apparatus and resealed with a quartz viewport using a copper gasket before performing each measurement. As described further below, the sides of the housing contain beryllium X-ray windows, and the front of the housing has several feedthroughs for electrical connections to the electrochemical cell, the heater, and the thermocouple used to monitor the sample temperature, as well as gas inlet and outlet ports. The entire apparatus fits between the X-ray optics in the Rigaku SmartLab X-ray diffractometer, as shown in Figure 2C.

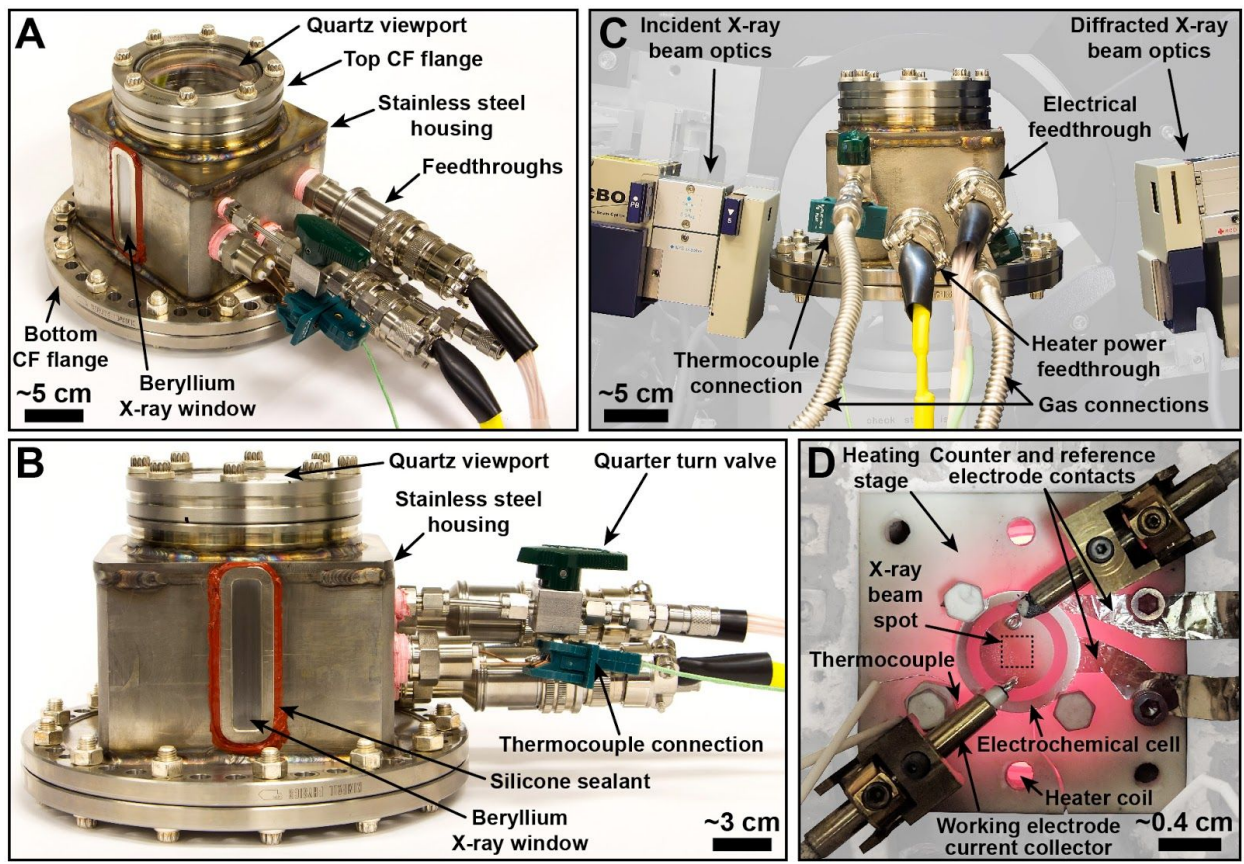

**Figure 2. A.** and **B.** Photographs of the exterior of the apparatus with key components labeled. **C.** Photograph of the apparatus mounted on the stage of the Rigaku SmartLab X-ray diffractometer. **D.** Top-down photograph of the electrochemical cell and heating stage taken through the top quartz viewport during electrochemical operation under a humidified $H_2$ atmosphere at 700 °C. In panels C and D, the contrast and brightness of the background have been adjusted to highlight the most important elements.

Beryllium X-ray windows (Heraeus Deutschland GmbH & Co. KG) for the incident and diffracted X-ray beams are attached to the left and right sides of the housing. Each window is 6.0 cm tall, 8.6 mm wide, and 0.25 mm thick. The size and locations of these windows enable the X-rays to impinge directly upon the center of the working electrode in the electrochemical cell, and permit access to the desired XRD 2θ scan range of 5 - 90°. Each window is fixed into a stainless steel support frame with silver solder. The frame is which is laser welded over a slot cut into the stainless steel housing (Leading Edge Fabricating, Inc.), and the welded edge is further coated with a high temperature silicone sealant (McMaster-Carr) to make the windows gas tight.

Gas inlet and outlet ports (Swagelok SS-200-1-2BT) are attached to the NPT tapped holes on the front of the housing. Each port is fitted with a quarter turn plug valve and a miniature quick-connect fitting (Swagelok SS-2P4T and SS-Q2-S-200). These fittings enable easy connections to an external manifold of valves, flow controllers, and a water bubbler to humidify the gas stream and regulate the gas flow rate into the apparatus. All internal components are constructed using materials that are resistant to degradation at high temperatures under inert or reducing gas atmospheres. The chamber is leak tight, as confirmed using a $H_2$ gas detector, which measures a $H_2$ concentration < 1 ppm a few cm away from all external surfaces of the apparatus at all times during operation. The internal volume of the apparatus is approximately 1 L, so at gas flow rates of 200 - 300 sccm, it is possible to completely exchange the internal gas within 15 - 20 min.

Inside the apparatus, the electrochemical cell is mounted to a ceramic heating stage, and electrical connections are made using various platinum contacts, as shown in Figure 2D. The heating stage is made from a Macor machinable ceramic sheet (Corning) with dimensions of 2.1 in × 2.1 in × 0.188 in. Several tapped holes in the ceramic stage enable the attachment of ceramic screws to hold down the electrochemical cell, thermocouple, and electrical contacts, as shown in Supporting Information Figure S1. The apparatus is designed for routine operation with a 20 mm diameter electrochemical cell, like the one pictured in Figure 1, but has the flexibility to accommodate diameters from 4 to 30 mm with minor changes to the Pt contact configuration. Figure 2D shows an electrochemical cell fastened to the Macor heating stage using two alumina screws.

As shown in Supporting Information S2, the electrical connections to the anode and reference electrode on the bottom of the electrochemical cell are made using Pt foils with Pt mesh on top to enable electrical contact while still permitting gas access to the electrodes. The Pt foil/mesh contacts are secured onto the ceramic heating stage using ceramic screws, and are cut to fit the shape of the anode and reference electrodes used in the example electrochemical cell pictured in Figure 1. These connections can easily be modified or replaced to accommodate other electrochemical cell geometries. Electrical connections to the working electrode on the top of the electrochemical cell are made using specially designed current collectors, which are pictured in Figure 2D and Supporting Information Figure S3. The current collectors consist of a Pt wire sealed into concentric ceramic and stainless steel supporting tubes. The Pt contact in the support tube can swivel on a trunnion in current collector bracket, which is screwed into the ceramic heating stage. A set screw in the current collector bracket is used to force the Pt wire down into contact with the cathode. The length and angle of these current collectors can easily be changed to accommodate different sample geometries. In Figure 2D, two current collectors are used to make redundant connections to the cathode to minimize potential variation due to sheet resistance across the thin film electrode. The current collectors are positioned near the edges of the cathode to prevent them from blocking the X-ray beam.

To connect the electrodes to the outside of the apparatus, all the Pt foil/mesh contacts and current collectors are attached to Pt wires which run through insulating ceramic tubes. The ends of these wires are crimped onto small stainless steel cylinders, which are pressure fit over the alumel pins in an NPT feedthrough which passes through on the front of the housing (MPF Products, Inc. A0198-12-NPT). The feedthrough can accommodate up to 10 independent electrical connections, but routinely four electrical connections are made (one to the anode, one to the reference, and two to the cathode). High temperature teflon tape (McMaster-Carr 44945K34, max temperature of 288 °C) is used on all NPT feedthrough junctions on the external housing side. On the outside of the apparatus, the electrical feedthrough is terminated with BNC cables used to connect to a potentiostat (Bio-Logic) for measuring the performance of the electrochemical cell.

The electrochemical cell is heated using a resistive spiral microheater (Micropyretics Heaters International Inc. MC-GAXP-130) underneath the ceramic stage. The electronic circuit controlling the heater is shown in Figure 3. Each end of the heater is connected to three Cu wires, each of which connects to one alumel pin on a six-pin NPT feedthrough which passes through the front of the housing (MPF Products, Inc. A0198-8-NPT). It is necessary to split the heater coil current across three pins to conform to the amperage limits of the feedthrough. Outside of the apparatus, the three pins connected to each heater coil terminal are shorted together again, and these leads connect to a transformer (Acme Electric TB181150) that supplies 24 VAC via a 120 VAC primary voltage source. The temperature of the electrochemical cell is measured and controlled using an S-type thermocouple (Omega) secured onto the heating stage with a ceramic screw approximately 2 mm from the edge of the electrochemical cell. The thermocouple is connected to a feedthrough (MPF Products, Inc. A0401-2-NPT) that passes through the front of the housing to enable connections outside the apparatus. The spiral microheater power supply and thermocouple are both connected to a process control system (Omega CNI-CB120) modified with a temperature controller (Dwyer 16B-23) that interfaces with controller software (Omega CN7-A) capable of setting a heating program and recording temperature data. The temperature controller operates two relays (Omega SSRL240DC25 and C25CNB130A, included in the CNI-CB120 process control system unit), as shown in detail in Figure 3. A 25 A and 1 A fuse (Omega ASK-4-1 and ASK-2S) protect the power and control circuits, respectively. All of the interior walls of the apparatus are lined with 0.188 in thick silicate ceramic insulation (McMaster-Carr 6841K1) to prevent heat loss through the stainless steel housing. The external surfaces of the apparatus reaches a temperature of 150 °C for the steel housing and 215 °C for the quartz window (not insulated) when the electrochemical cell temperature is at its maximum value (745 °C).

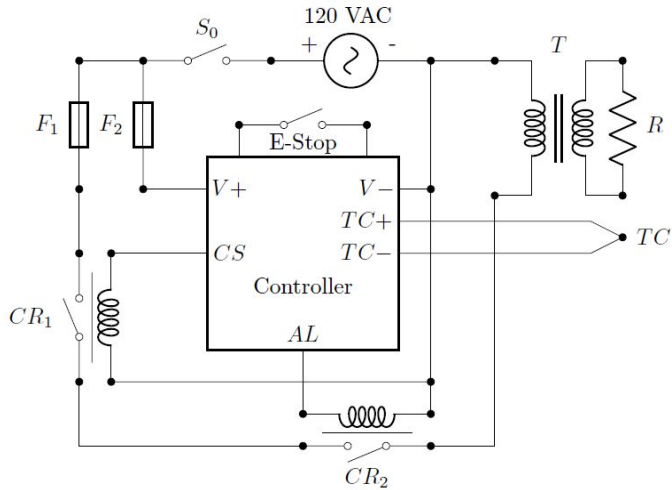

**Figure 3.** Electrical diagram for temperature control consisting of a process controller system with a modified temperature controller, transformer, and spiral microheater. The circuit is fed by a standard 120 VAC source. A temperature controller operates a relay ($CR_1$) with a control signal (*CS*) to run power to a transformer (*T*), which in turn downregulates the 120 VAC primary power line to 24 VAC. The 24 VAC is fed into the microheater (*R*) that heats the Macor plate shown in Figure 2. An alarm control signal (*AL*) from the controller can signal to another relay ($CR_2$) to shut off power when the temperature measured by the S-type thermocouple (*TC*) connected to the controller's sensor ports (*TC+* and *TC-*) exceeds that of the controller setpoint temperature. Two fuses $F_1$ and $F_2$ protect the source and controller power lines, respectively. The controller is also equipped with an emergency stop (E-Stop) and an on/off switch ($S_0$).

# IV. Example *Operando* X-ray Diffraction Measurements

We performed an example measurement using the electrochemical cell shown in Figure 1 to demonstrate the capabilities of our *operando* XRD apparatus. Figure 4A shows an XRD spectrum of the electrochemical cell inside the apparatus at room temperature. The spectrum reveals numerous reflections arising from the Pd cathode and the BZCY electrolyte. These data demonstrate that the apparatus enables the collection of XRD spectra in the desired range of 5 - 90°.

Figure 4B shows 69 XRD spectra of the Pd (111) reflection collected every 10 minutes during electrochemical cell operation. The color of each trace corresponds to the average temperature during the XRD spectrum collection time. The first 13 spectra were collected while the cell was held at 745 ± 2 °C for 2.1 hr. Then the electrochemical cell was cooled at 1.3 °C/min to a final temperature of 37 °C. As the temperature of the cell changes, shifts in the position of the Pd (111) reflection are observed. Between 745 and 188 °C, the reflection shifts to higher 2θ position due to thermal contraction of the cathode upon cooling. Below 188 °C, the (111) reflection shifts to lower 2θ position due to lattice expansion caused by spontaneous absorption of H into the Pd cathode from gas phase $H_2$. As the temperature continues to decrease, the cathode eventually undergoes a phase change to form *β*-PdH$_x$, which causes the lattice to expand [12]. These data demonstrate that the apparatus can detect changes in crystal structure of the operating electrochemical cell.

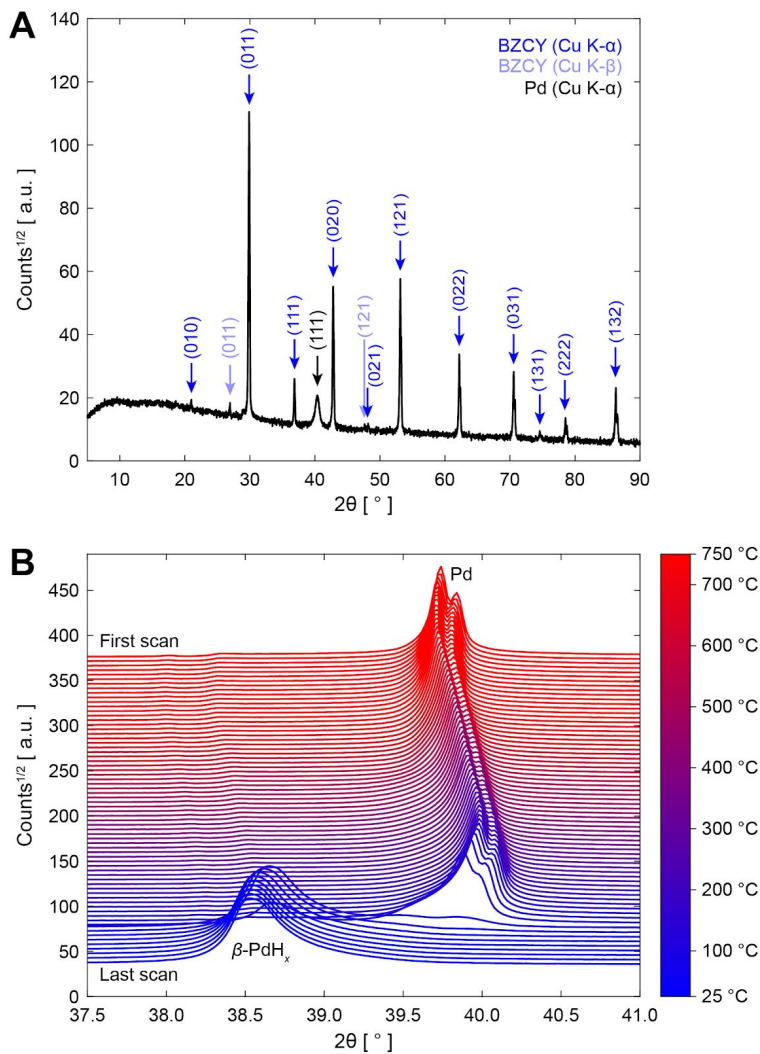

**Figure 4. A.** XRD spectrum of the electrochemical cell inside the apparatus at room temperature before heating and beginning electrochemical operation, demonstrating the capability our apparatus to collect spectra in the range of 5 - 90°. Many reflections arising from the BZCY electrolyte are observed, as expected. A single reflection corresponding to the (111) plane of the Pd cathode is observed, indicating that the Pd film posses significant preferred orientation in the (111) direction. **B.** XRD spectra of the Pd/PdH$_x$ cathode (111) reflection collected every 10 min during electrochemical operation. The temperature was held at 745 ± 2 °C to begin, then the electrochemical cell was cooled at 1.3 °C/min to a final temperature of 37 °C. Each spectrum is colored according to the average temperature during the scan time, and the spectra are vertically offset for clarity. The asymmetry in the peak profile observed at high temperatures arises from the splitting between the Cu K-α1 and K-α2 X-ray wavelengths.

Figure 5 shows electrochemistry, temperature, and structural data from the operating electrochemical cell as a function of experimental time (bottom horizontal axis) and the corresponding XRD scan number (top horizontal axis). These data come from the same experiment as Figure 4. The top two panels show the instantaneous and average current density ($J$) over each XRD scan, followed by the instantaneous and average cathode potential ($E_{cat}$) with respect to the reversible hydrogen electrode reference potential ($E_{RHE}$). The third panel from the top shows the instantaneous and average temperature during each XRD scan ($T$). Finally, the bottom panel shows the Pd/PdH$_x$ lattice parameter extracted from the position of the (111) reflection for each spectrum shown in Figure 4B. These data indicate that the applied current does not cause changes in the cathode structure. Nevertheless, the *operando* XRD apparatus is capable of simultaneously recording electrochemical performance data and XRD spectra at temperatures up to 745 °C, which exceeds the design objective of 725 °C.

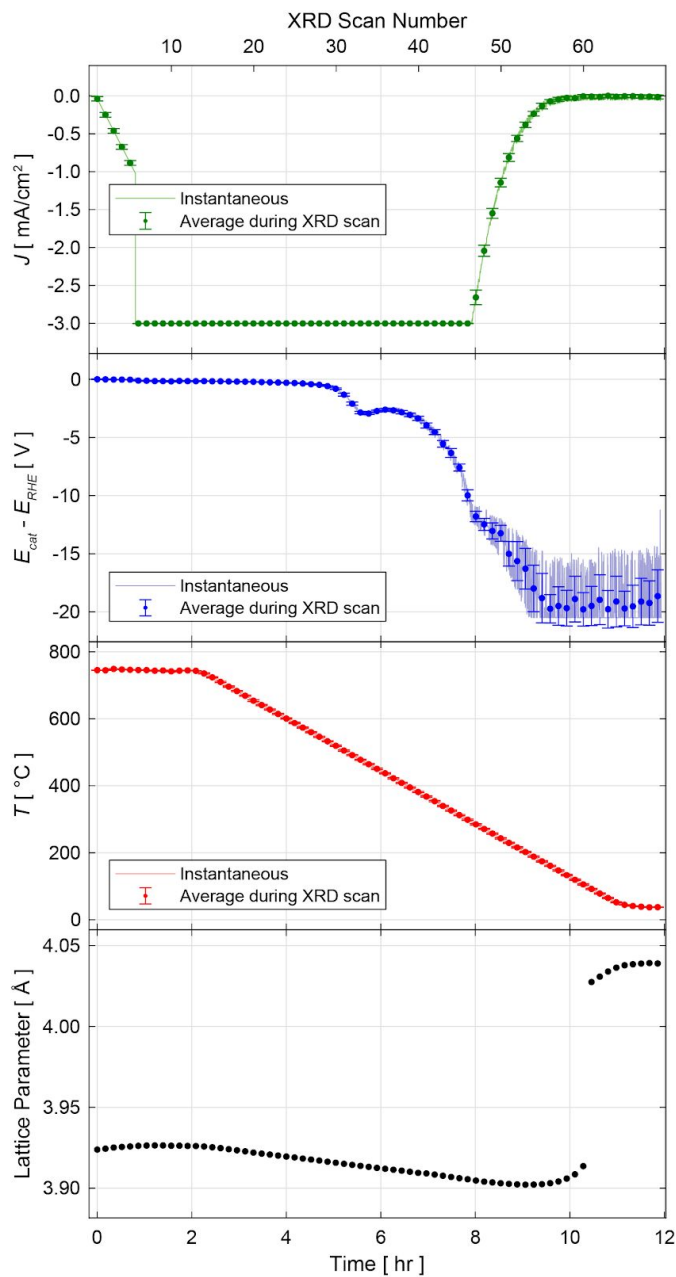

**Figure 5.** Electrochemistry, temperature, and cathode structural data for the electrochemical cell during high temperature electrochemical operation. The data in these plots are from the same set of measurements displayed in Figure 4. The top two panels show the applied current density ($J$) and corresponding cathode potential ($E_{cat}$) versus the reversible hydrogen electrode potential ($E_{RHE}$) over time. The third panel from the top displays the temperature ($T$). The error bars in the top three panels each correspond to one standard deviation in the measured value of each parameter during the collection time for each XRD spectrum. The bottom panel displays the lattice parameter of the Pd/PdH$_x$ cathode measured from the XRD spectra.

# V. Extensions to Measurement Capabilities

The capabilities of our apparatus could be extended to study other conditions or enhance performance with some modifications. Increasing the height of the beryllium X-ray windows could enable access to at 2θ angle of 0°. While this is not necessary to collect XRD spectra to analyze the crystal structure of electrochemical cell materials, it could be useful for sample height alignment in the diffractometer.

The maximum operating temperature of the apparatus could likely be increased to exceed the maximum demonstrated operating temperature of 745 °C by adding extra insulation to the interior walls of the apparatus to further reduce heat loss through the stainless steel housing. It is not possible to measure the temperature uniformity of our apparatus at present, but there is likely some variation in the local temperature at different positions across the electrochemical cell. Using a larger heater coil could could provide improved spatial uniformity of the temperature across the electrochemical cell, and also potentially contribute to higher maximum temperatures.

Our apparatus could be modified to be resistant to oxidizing gas atmospheres such as air. This would enable *operando* XRD measurements of oxygen electrodes in SOFCs and SOECs. Presently, the primary component of our apparatus that would corrode at high temperatures in air is the copper wires that connect the heater coil to the heater power feedthrough. These wires could be replaced with a different material such as Pt, which is resistant to oxidation. Some other apparatus components, such as the stainless steel current collector brackets, could also oxidize, but this would not significantly inhibit the performance of the apparatus.

Finally, this apparatus could be used to investigate structural changes in other high temperature electrically active systems beyond solid oxide electrochemical cells. For example, with only minor modifications to the electrical contacts, the apparatus could be used to study high temperature piezoelectric materials [14, 15]. With some more extensive modifications to the ceramic heater stage to accommodate liquid samples, the apparatus could also be used for simultaneous electrical and phase measurements on liquid crystalline materials [16, 17, 18].

# VI. Conclusion

Our apparatus provides the first experimental capability to measure structural changes in fuel electrodes of high temperature solid oxide electrochemical cells under reducing atmospheres via *operando* XRD. The apparatus is compatible with a conventional laboratory powder X-ray diffractometer, which enables high throughput measurements. Solid oxide electrochemical cells with diameters ranging from 4 - 30 mm can be tested in the apparatus. Rapid setup of these electrochemical cells is facilitated by flexible electrical connections. The apparatus can make up to four electrical connections to the electrochemical cell, enabling experiments with both two-electrode and three-electrode configurations. Electrochemical cells can be tested at temperatures up to 745 °C under reducing (e.g. dry or humidified $H_2$) or inert (e.g. Ar or $N_2$) gas atmospheres. As a result of this unique combination of capabilities, this apparatus will enable experiments to assess the effect of operating conditions on electrode and electrolyte structure, making this an important tool for identify strategies to limit degradation and create durable, high performance solid oxide electrochemical cells.

# Acknowledgments

The authors would like to acknowledge Dr. Charles Settens for assistance with apparatus design, Dr. James Hunter for assistance with apparatus design and fabrication, and Dr. David Fork, Dr. Ross Koningstein, and Matt Trevithick for helpful discussions. The XRD measurements were performed at the MIT Center for Materials Science and Engineering, a MRSEC Shared Experimental Facility supported by the National Science Foundation under award number DMR-14-19807. Financial support was provided by Google LLC. This material is based upon work supported by the National Science Foundation Graduate Research Fellowship under Grant No. 1122374.

# Supporting Information

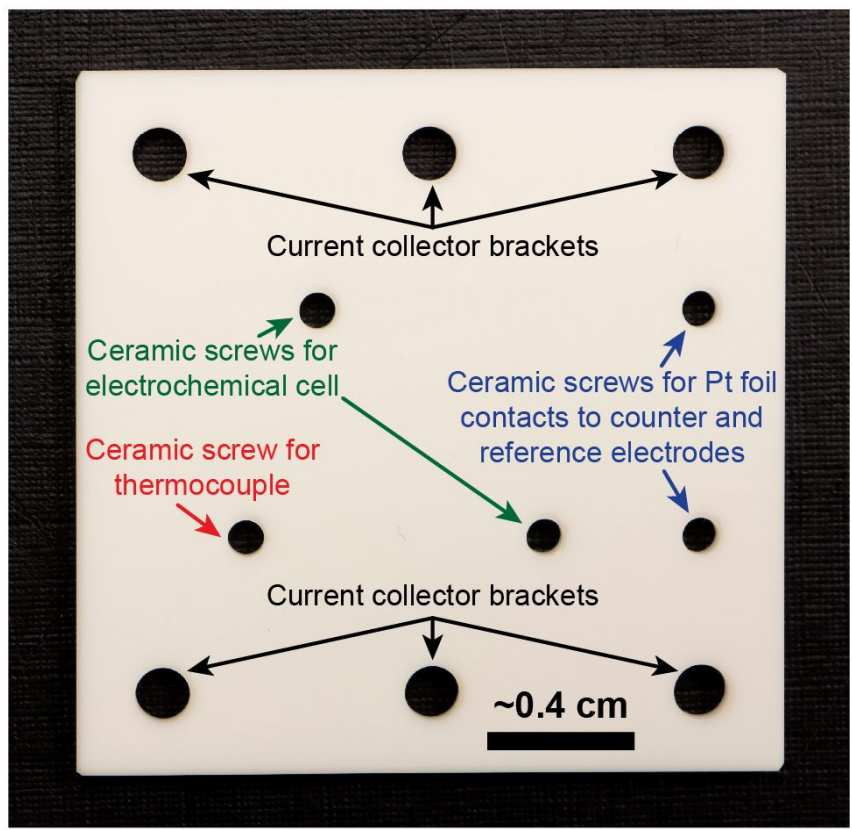

**Figure S1.** Photographs of Macor ceramic heating stage with tapped holes used to secure the electrochemical cell, thermocouple, and Pt electrical contacts.

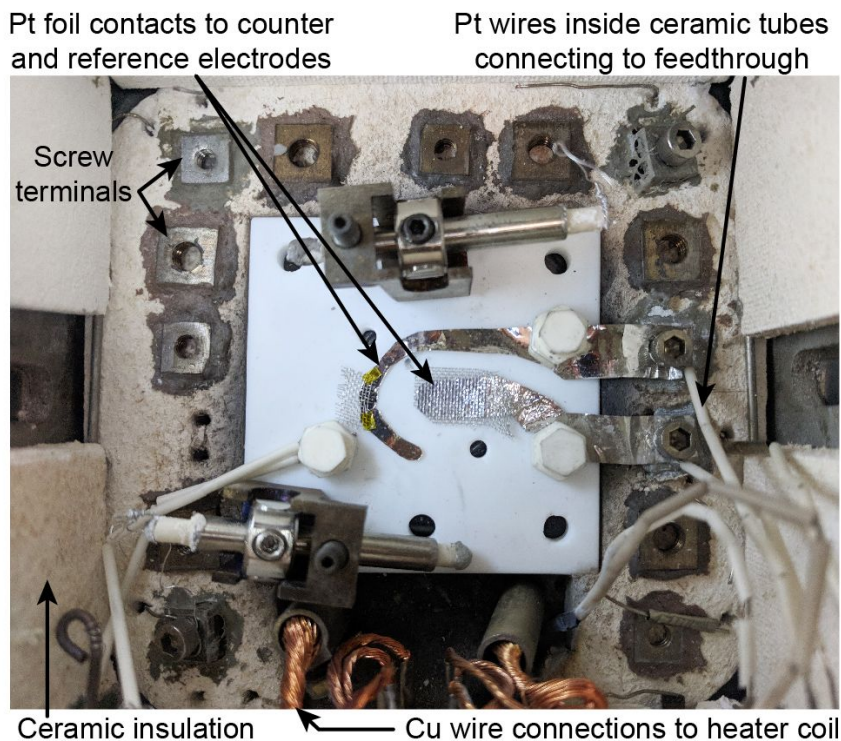

**Figure S2.** Photograph of the interior of the apparatus. The Pt foil contacts to the reference and counter electrodes are specifically shaped to match the example electrochemical cell used in this study, but these connections can easily be modified or replaced to accommodate other cell geometries. Stainless steel square nuts bonded into the ceramic insulation around the ceramic heating stage are used as screw terminals for making electrical connections from the Pt foils and current collectors to the ceramic tube-insulated wires that connect to the feedthroughs. The large number of these screw terminals provides flexibility for modifying the electrical connections as necessary.

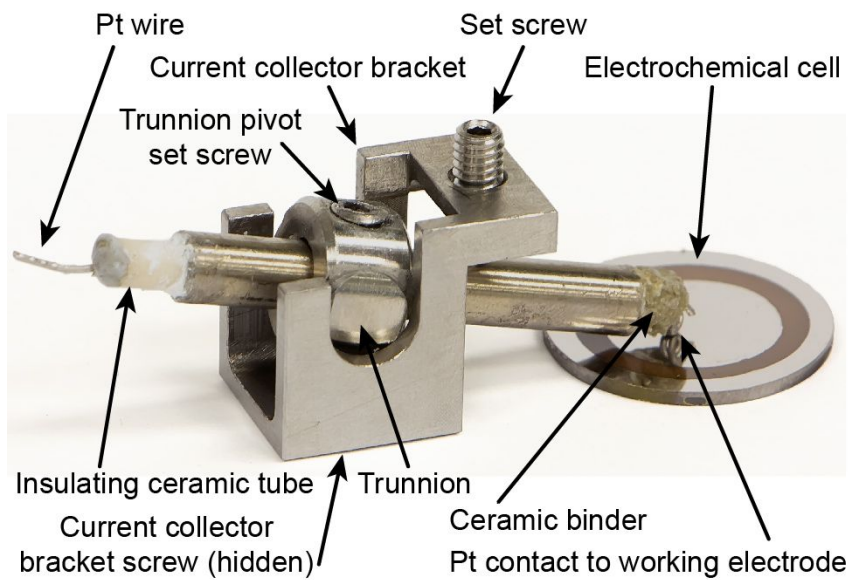

**Figure S3.** Photograph of working electrode current collector.

# References

1. Z. Gao, L.V. Mogni, E.C. Miller, J.G. Railsback, and S.A. Barnett. "A perspective on low-temperature solid oxide fuel cells." Energy & Environmental Science, 2016. 9 (5): 1602-1644. http://dx.doi.org/10.1039/c5ee03858h
2. M.L. Traulsen, C. Chatzichristodoulou, K.V. Hansen, L.T. Kuhn, P. Holtappels, and M.B. Mogensen. "Need for In Operando Characterization of Electrochemical Interface Features." ECS Transactions, 2015. 66 (2): 3-20. http://dx.doi.org/10.1149/06602.0003ecst
3. S. Wolf, N.A. Canas, and K.A. Friedrich. "In Situ X-Ray Diffraction and Stress Analysis of Solid Oxide Fuel Cells." Fuel Cells, 2013. 13 (3): 404-409. http://dx.doi.org/10.1002/fuce.201300025
4. H. Orui, R. Chiba, K. Nozawa, H. Arai, and R. Kanno. "High-temperature stability of alumina containing nickel-zirconia cermets for solid oxide fuel cell anodes." Journal of Power Sources, 2013. 238: 74-80. http://dx.doi.org/10.1016/j.jpowsour.2013.03.056
5. J. Jeong, A.K. Azad, H. Schlegl, B. Kim, S.-W. Baek, K. Kim, H. Kang, and J.H. Kim. "Structural, thermal and electrical conductivity characteristics of $Ln_{0.5}Sr_{0.5}Ti_{0.5}Mn_{0.5}O_{3\pm d}$ (Ln: La, Nd and Sm) complex perovskites as anode materials for solid oxide fuel cell." Journal of Solid State Chemistry, 2015. 226: 154-163. https://doi.org/10.1016/j.jssc.2015.02.001
6. J.S. Hardy, J.W. Templeton, D.J. Edwards, Z.G. Lu, and J.W. Stevenson. "Lattice expansion of LSCF-6428 cathodes measured by in situ XRD during SOFC operation." Journal of Power Sources, 2012. 198: 76-82. http://dx.doi.org/10.1016/j.jpowsour.2011.09.099
7. J.S. Hardy, C.A. Coyle, J.F. Bonnett, J.W. Templeton, N.L. Canfield, D.J. Edwards, S.M. Mahserejian, L. Ge, B.J. Ingram, and J.W. Stevenson. "Evaluation of cation migration in lanthanum strontium cobalt ferrite solid oxide fuel cell cathodes via in-operando X-ray diffraction." Journal of Materials Chemistry A, 2018. 6 (4): 1787-1801. http://dx.doi.org/10.1039/c7ta06856e
8. I. Kivi, J. Aruvali, K. Kirsimae, A. Heinsaar, G. Nurk, and E. Lust. "Changes in LSC and LSCF Cathode Crystallographic Parameters Measured by Electrochemical in situ High-Temperature XRD," in "Solid Oxide Fuel Cells 13." T. Kawada and S.C. Singhal, Editors. 2013. 1841-1849. http://dx.doi.org/10.1149/05701.1841ecst
9. S. Volkov, V. Vonk, N. Khorshidi, D. Franz, M. Kubicek, V. Kilic, R. Felici, T.M. Huber, E. Navickas, G.M. Rupp, J. Fleig, and A. Stierle. "Operando X-ray Investigation of Electrode/Electrolyte Interfaces in Model Solid Oxide Fuel Cells." Chemistry of Materials, 2016. 28 (11): 3727-3733. http://dx.doi.org/10.1021/acs.chemmater.6b00351
10. L. Zhao, B.B. He, B. Lin, H.P. Ding, S.L. Wang, Y.H. Ling, R.R. Peng, G.Y. Meng, and X.Q. Liu. "High performance of proton-conducting solid oxide fuel cell with a layered $PrBaCo_2O_{5+delta}$ cathode." Journal of Power Sources, 2009. 194 (2): 835-837. http://dx.doi.org/10.1016/j.jpowsour.2009.06.010

11. L. Yang, C.D. Zuo, S.Z. Wang, Z. Cheng, and M.L. Liu. "A novel composite cathode for low-temperature SOFCs based on oxide proton conductors." Advanced Materials, 2008. 20 (17): 3280-+. http://dx.doi.org/10.1002/adma.200702762
12. Manchester, F. D., San-Martin, A., & Pitre, J. M. (1994). The H-Pd (hydrogen-palladium) system. Journal of phase equilibria, 15(1), 62-83. https://doi.org/10.1007/BF02667685
13. A. Czerwinski, I. Kiersztyn, M. Grden, and J. Czapla. "The study of hydrogen sorption in palladium limited volume electrodes (Pd-LVE) I. Acidic solutions." Journal of Electroanalytical Chemistry, 1999. 471 (2): 190-195. http://dx.doi.org/10.1016/s0022-0728(99)00276-4
14. D. Damjanovic. "Materials for high temperature piezoelectric transducers." Current Opinion in Solid State & Materials Science, 1998. 3 (5): 469-473. http://dx.doi.org/10.1016/S1359-0286(98)80009-0
15. X. Jiang, K. Kim, S. Zhang, J. Johnson, and G. Salazar. "High-temperature piezoelectric sensing." Sensors, 2013. 14 (1): 144-169. http://dx.doi.org/10.3390/s140100144
16. S. Singh. "Phase transitions in liquid crystals." Physics Reports, 2000. 324 (2): 107-269. http://dx.doi.org/10.1016/S0370-1573(99)00049-6
17. "Structural Studies of Liquid Crystals by X-Ray Diffraction," in "Handbook of Liquid Crystals Set." http://dx.doi.org/10.1002/9783527619276.ch8ca
18. V. Hinkov, D. Haug, B. Fauqué, P. Bourges, Y. Sidis, A. Ivanov, C. Bernhard, C.T. Lin, and B. Keimer. "Electronic Liquid Crystal State in the High-Temperature Superconductor YBa2Cu3O6.45." Science, 2008. 319 (5863): 597-600. http://dx.doi.org/10.1126/science.1152309